\documentclass[prd,aps]{revtex4}
\usepackage{graphicx}
\usepackage{amsmath}
\usepackage{amsfonts}
\usepackage{amssymb}
\usepackage{cancel}
\DeclareGraphicsExtensions{.pdf,.png,.jpg}

\begin{document}
\preprint{{hep-th/yymmnnn} \hfill {UCVFC-DF-17-2005}}
\title{Mimetic discretization of the Abelian Chern-Simons theory and link invariants.}
\author{ Cayetano Di Bartolo$^{1}$, Javier Grau$^{1}$ and Lorenzo Leal$^{1,2}$}

\affiliation {1. Departamento de F\'{\i}sica, Universidad Sim\'on Bol\'{\i}var, Apartado Postal 89000,
Caracas 1080-A, Venezuela.\\2. Centro de F\'{\i}sica Te\'{o}rica y
Computacional, Facultad de Ciencias, Universidad Central de
Venezuela, Apartado Postal 47270, Caracas 1041-A, Venezuela. \\ }

\begin{abstract}
A mimetic discretization of the Abelian Chern-Simons theory is presented. The study relies on the formulation of a theory of differential forms in the lattice, including a consistent definition of the Hodge duality operation. Explicit expressions for the Gauss Linking Number in the lattice, which correspond to their continuum counterparts are given. A discussion of the discretization of  metric structures in the space of transverse vector densities is presented. The study of these metrics could serve to obtain explicit formulae for knot an link invariants in the lattice.
\end{abstract}

\maketitle

\section{Introduction}

The relationship between quantum Chern-Simons theory and knot theory is an active field of research both in physics and mathematics \cite{Witten,Labastida}. The key contribution in this field, due to E. Witten, stated that the vacuum expectation value of the Wilson loop in the Chern-Simons theory yields knot invariants related with the Jones polynomial \cite{Witten,Jones}.

Besides their interest from the mathematical side, knot invariants play an important role in the loop representation of quantum gravity \cite{Rov-Smo,Gambini-91} based in Ashtekar variables\cite{Ashtekar}. In this formulation,  wave functionals are just knot-invariants, hence, it is useful to have analytical expressions for knot-invariants that can be properly handled through the machinery of loop-calculus. A systematic way to obtain expressions of this kind consists on considering the perturbative expansion of the Wilson loop in the quantum Chern-Simons theory \cite{Guadagnini,Gae-93}. Under very general conditions, it can be shown that each order in this perturbative expansion is a knot invariant. Hence, following this method one can get, in principle, an infinity of such invariants. It should be said that there are also "knot" and "link" invariants associated to the spin-networks formulation of quantum gravity \cite{Rovelli}, which is the modern setting in which loop quantum gravity is studied.

There also exists another method to obtain knot-invariants, which deals with the classical field formulation. This method consists on considering the on-shell action of a diffeomorphism-invariant theory coupled to external sources with support on a curve (or several curves, if one desires to consider link-invariants instead of knot-invariants). The  metric independence of the theory guarantees the topological character of the on-shell action, which yields analytical expressions for knots (or link) invariants. This method was first considered for abelian theories \cite{Lorenzo-92}, and then extended for non-abelian ones \cite{Lorenzo-02}.

Regardless one uses the quantum or the classical method, the link-invariants (i.e., those invariants involving several loops) that rise from the Chern-Simons theory are well defined (provided that the curves involved do not intersect each other). This is not always the case for knot-invariants, which can suffer from ambiguities (ref), as the Gauss self-linking number exemplifies \cite{rolfsen}. A way to cure these ambiguities consists on  choosing  a "framing": instead of considering a curve, one takes a strip whose width will be taken to vanish after the entanglement properties are calculated. Another way of handling the ambiguities is  through  the discretization of the knot or link invariants, since the lattice can provide a natural framing. On the other hand, a discretization of topological field theories or loop quantum gravity can be also useful for calculational purposes.

There are many ways to discretize a physical theory, since there are many forms to convert differential equations into finite difference equations, or to set a sensible discretization of vectorial operators such as curl, gradient and divergence. The line, surface and volume integral of vectorial fields admit different discretizations as well. When the discretization is made in such a way that most of the significative mathematical properties (in the continuum) of the objects involved are preserved in the lattice, one says that it is "mimetic". There is a vast literature \cite{stanly} about mimetic discretizations in the framework of classical mechanics and electromagnetism. Also, the Hamiltonian approach of QCD in the lattice \cite{Kogut} implicitly deals with a mimetic discretization of the Yang-Mills theory, although this fact is seldom stressed.

 Some years ago, a "consistent discretization" of the linearized theory of General Relativity, both in the traditional metric variables and in the Ashtekar variables was considered, yielding a mimetic discretization of the theory  \cite{Gae-04}. A mimetic discretization of the Abelian  Chern-Simons theory in a lattice was proposed by Fröhlich and Marchetti \cite{Marchetti} and studied by several authors \cite{varios}. A mayor problem related to the discretization of theories with differential forms consists on the mimetization of the Hodge dual operation. In reference \cite{Gambini-97} another mimetic discretization of the Chern-Simons theory was considered, and using the quantum method explained above the authors obtained a formula for the self-linking number in the lattice. There was argued that the framing problem
is related to the discretization of the Hodge dual operation. Hence, a prescription for this operation was  given in that reference, but it is knot completely mimetic, since taking the dual twice  does not lead back to the starting form.

In this article we study a mimetic discretization of the Abelian Chern-Simons theory and discuss how to obtain knot invariants in the lattice from the evaluation of the discrete on-shell action. This is achieved by means of the introduction of a formalism for the  mimetic discretization of differential forms in the three-dimensional cubic lattice, which includes a natural formulation of  Hodge duality.  We write down the Chern-Simons action coupled to external  currents associated to curves in the lattice (which are also mimetically discretized), obtain the equations of motion, solve them, and calculate the action on shell, which produces formulae for knot (or link) invariants in the lattice.  Also, we address (to some extent) the problem of taking the analytical expressions for some higher order knot and link invariants already known in the continuum, and producing discrete formulae for them in a mimetic form. To this end, we elaborate further on the study of the "metrics" in the space of transverse functions studied in reference \cite{Cayetano}. These metrics can be seen as building blocks in the construction of invariants, both in the continuum and in the lattice.

\section{Differential forms in the cubic lattice}

We shall denote  by $v$, $l$ and $p$, the sites, oriented links and oriented plaquettes of the cubic lattice, respectively, . $\overline{l}$  ($\overline{p}$) denotes the link (plaquette) that coincides with $l$ ($p$) but has opposite sense. The elementary cubes, plaquettes, links and sites are the the $q$-cells of the lattice, with $q=3,2,1,0$ respectivelly. $\mathbb{L}$ means the distance between nearest neighboring sites. Sometimes it will be convenient to label  the $q$-cells using sites and unitary vectors as follows. Given a right handed orthonormal basis $\{\hat{e}_1, \hat{e}_2, \hat{e}_3\}$ formed by vectors parallel to the links of the lattice, we can write, for instance,  $v' = v +\mathbb{L} \hat{e}$, meaning that the sites  $v'$ and  $v$ are nearest neighbors separated by a link in the direction of $\hat{e}$. We shall also use the notation

\begin{equation}
\mathbb{L}_i \equiv \mathbb{L} \hat{e}_i \quad\text{and}\quad \mathbb{L}_{123} \equiv \mathbb{L}(\hat{e}_1 +\hat{e}_2 +\hat{e}_3)\,.
\end{equation}

 If the link $l$ starts at $v$ and ends at $v'= v +\mathbb{L}\,\hat{e}$, we shall also set $l = L(v,\hat{e})$. With this notation, the opposite link is $\bar{l} = L(v',-\hat{e})$. The site where $l$ starts will be denoted as  $.l$  and the final one will be $l\bm{\cdot}$. We say that a link $l$ pointing along one of the basis vectors $\hat{e}_i$ is a positive one, and we  write $l>0$.
Consider a plaquette coplanar with $\hat{e}_i$ and $\hat{e}_j$, whose orientation is given by $\hat{e}_i\times\hat{e}_j$. Let $v$ the site of  $p$ having lower values of its coordinates. Then we say that $p= P(v,\hat{e}_i,\hat{e}_j)$. A plaquette whose normal vector points along some of the $\hat{e}_i$ is a positive one, and we write $p>0$.

A discrete 0-form is a function taking values on the sites of the lattice. If $\lambda$ is a 0-form, $\lambda_v$ is the value it takes at site $v$. The domain of a  discrete 1-form, instead, is the set of links of the lattice. Let  $A$ be a 1-form and $l=L(v,\hat{e}_i)$ a link; the evaluation of $A$ at $l$ will be indistinctly denoted by $A_l=A_i(v)$. We shall also say that $A_i(v)$ is the component along $\hat{e}_i$ of the vector field $A$ evaluated at $v$. We ask the 1-forms to obey

\begin{equation}
A_{\overline{l}} = - A_l \,,
\end{equation}
which  is the lattice version of the "continuum" equation $\vec{A}\cdot(-\hat{e}_i) = -\vec{A}\cdot(\hat{e}_i)$.
A discrete 2-form takes values on the plaquettes: $B_p=B_{i\,j}(v)$ means the evaluation of the 2-form $B$ on the plaquette $p=P(v,\hat{e}_i, \hat{e}_j)$. We impose the condition

\begin{equation}
B_{\overline{p}} = - B_p \quad\text{or equivalently}\quad B_{j\,i}(v) = -B_{i\,j}(v)\,.
\end{equation}
Also, we set  $B_{i\,i}(v)\equiv 0$. As in the continuum, discrete 2-forms are antisymmetric.
Finally, we define a 3-form as a function taking values at the cubes $c$  of the lattice. If $\Gamma$ is a  3-form, we write
\begin{equation}
\Gamma_c = \Gamma_{1\,2\,3}(v) = \Gamma_{i\,j\,k}(v) = - \Gamma_{k\,j\,i}(v)
\end{equation}
Here, $v$ is the site with lower coordinate values belonging to $c$, and $\{i,j,k\}$ is a cyclic permutation of $\{1,2,3\}$. We also set $\Gamma_{i\,j\,k}(v)\equiv 0$ whenever any pair of indexes take the same value.

Given a 0-form $f(v)$, we define its discrete partial derivative in the direction of  $\hat{e}_i$ as
\begin{equation}
\partial_i f(v) \equiv f(v+\mathbb{L}_i) - f(v)\,.
\end{equation}
It is easy to see that
\begin{equation}
\partial_j \partial_i f(v) = \partial_i \partial_j f(v)\,,
\end{equation}
and that when the lattice spacing tends to zero, the discrete derivative tends to the continuum derivative.

We define a $q$-form $\phi$ as a function over the $q$-cells. Its  exterior derivative  $(d \phi)$ is a $(q+1)$-form defined by
\begin{equation}
(d \phi)_{(q+1)\text{-cell}} = \sum_{q\text{-cell} \in \partial(q+1)\text{-cell}} \phi_{q\text{-cell}} ,
\end{equation}
 where $\partial(q+1)\text{-cell}$ is the boundary of the $(q+1)$-cell.
 Let us see how this formula looks in detail. For a  0-form $\lambda$ we have
\begin{equation}
(d \lambda)_l = \lambda_{l\bm{\cdot}} - \lambda_{\bm{\cdot}l}\,\,  .
\end{equation}
If $l=L(v,\hat{e}_a)$ we have
\begin{equation}
(d \lambda)_{L(v,\hat{e}_a)} = \lambda(v+\mathbb{L}_a) - \lambda(v) = \partial_a \lambda(v)\,.
\end{equation}
Now, let $A$ be a  1-form. Its exterior derivative is given by
\begin{equation}
(d A)_p = \sum_{l\in p} A_l,
\end{equation}
 where the links $l$ belonging to the edge of  $p$ are taken accordingly to the right hand rule with respect to the orientation of $p$. If $p=P(v,\hat{e}_i,\hat{e}_j)$, then
\[
(d A)_{i\,j}(v) = A_i(v)-A_i(v+\mathbb{L}_j) + A_j(v+\mathbb{L}_i) - A_j(v)\,,
\]
hence
\begin{equation}
(d A)_{i\,j}(v) = \partial_i A_j(v) - \partial_j A_i(v)\,.
\end{equation}
Next, take $B$ to be a  2-form. We have
\begin{equation}
(d B)_c = \sum_{p\in c} B_p
\end{equation}
where the plaquettes $p$ that bound the cube $c$ are oriented outwards. If  $v$ is the site of $c$ having lesser values of their components, then
\begin{equation}
(d B)_c =  -B_{3\,1}(v)+ B_{3\,1}(v+\mathbb{L}_2) - B_{1\,2}(v) +B_{1\,2}(v+\mathbb{L}_3) \\
 - B_{2\,3}(v) + B_{2\,3}(v+\mathbb{L}_1) \\
=  \partial_2 B_{3\,1}(v) + \partial_3 B_{1\,2}(v) + \partial_1 B_{2\,3}(v)\,  .
\end{equation}
Therefore we have
\begin{equation}
(d B)_{i\,j\,k}(v) = \partial_{(i} B_{j\,k)_{\scriptstyle c}}(v) = \frac{1}{2}\delta^{d\,e\,f}_{i\,j\,k}\,\partial_d B_{e\,f}(v)\, ,
\end{equation}
where $\delta^{d\,e\,f}_{i\,j\,k}$ stands for the antisymmetrized product of three Kronecker's deltas.
Finally, the exterior derivative of a 3-form is taken to be zero. Summarizing, if $\phi$ is a $q$-form, its exterior derivative obeys
\begin{equation}
(d \phi)_{i_1 \cdots i_{q+1}}(v) = \frac{1}{q!} \delta_{i_1 \cdots i_{q+1}}^{j_1 \cdots j_{q+1}} \partial_{j_1} \phi_{j_2 \cdots j_{q+1}}(v)\, .
\end{equation}
It can be easily shown that
\begin{equation}
dd \phi= 0\,
\end{equation}
for any discrete form $\phi$. Thus, we have a mimetic  version of differential forms and exterior derivative in the cubic lattice.

There exist many ways for defining the Hodge dual $\# \omega$ of a form $\omega$ in the lattice. We shall define it by relating forms in the "direct lattice" $\mathbb{R}$ with forms in the "dual lattice" $\mathbb{R}^d$, which is defined, as usual, in such a way that its sites coincide with the centers of the cubes of the former one $\mathbb{R}$. Hence, links of the dual lattice go across plaquettes of the "direct" lattice, while plaquettes of the dual lattice are crossed by links of the "direct" one. Our total space will be  the union of both $\mathbb{R}$ and $\mathbb{R}^d$. If we want to stress that an elementary cell belongs to a lattice which is the dual one of another one, we shall attach the superscript  $*$.
 Given the  0-form $\lambda$, we define its Hodge dual $\# \lambda$ as the 3-form given by
\begin{equation}
(\# \lambda)_{c^*} = \lambda_v
\end{equation}
where $c^*$ is the cube whose center coincides with the site $v$  ($c^*$ belongs to the "dual lattice" if $v$ belongs to the "direct" one). The components $(\# \lambda)_{i\,j\,k}(v^*)$ satisfy
\begin{equation}
(\# \lambda)_{i\,j\,k}(v^*) = \varepsilon_{i\,j\,k}\lambda(v) \quad\text{with}\quad v^* = v - \frac{1}{2}\, \mathbb{L}_{123}\,.
\end{equation}

For a 1-form $A$, we define its  Hodge dual $\# A$ as the  2-form that satisfies
\begin{equation}
(\# A)_{p^*} = A_l
\end{equation}
 where $p^*$ is the plaquette dual to the link $l$. It can be seen that
\begin{equation}
(\# A)_{i\,j}(v^*) = \varepsilon_{i\,j\,k} \,A_k(v) \quad\text{with}\quad v^* = v +\frac{1}{2}\, \mathbb{L}_{123} - \mathbb{L}_i - \mathbb{L}_j\,.
\end{equation}
Now let $B$ be a 2-form. Its  Hodge dual $\# B$ is defined as the 1-form satisfying
\begin{equation}
(\# B)_{l^*} = B_p
\end{equation}
 where $l^*$ is the link dual to the plaquette $p$. We have
\begin{equation}
(\# B)_{i}(v^*) = \frac{1}{2} \varepsilon_{i\,j\,k} \,B_{j\,k}(v) \quad\text{with}\quad v^* = v +\frac{1}{2}\, \mathbb{L}_{123}- \mathbb{L}_i\,.
\end{equation}
Finally, given a 3-form $\Gamma$ we  define its Hodge dual $\# \Gamma$ to be the  0-form which fulfils
\begin{equation}
(\# \Gamma)_{v^*} = \Gamma_c
\end{equation}
with $v^*$ being the site dual to the cube $c$. One also has
\begin{equation}
(\# \Gamma)(v^*) = \frac{1}{3!} \varepsilon_{i\,j\,k}\, \Gamma(v)_{i\,j\,k} \quad\text{with}\quad v^* = v +\frac{1}{2}\, \mathbb{L}_{123}\,.
\end{equation}
These definitions and results may be summarized as follows. If $\phi$ is a $q$-form, its Hodge dual satisfies
\begin{subequations}
\begin{equation}
(\# \phi)_{i_1 \cdots i_{3-q}}(v^*) = \frac{1}{q!} \varepsilon_{i_1 \cdots i_{3-q}\, j_1 \cdots j_{q}} \phi_{j_1 \cdots j_{q}}(v)
\end{equation}
where
\begin{equation}
v^* = v +\frac{\mathbb{L}}{2}(\hat{e}_1 +\hat{e}_2 + \hat{e}_3 - 2  \hat{e}_{i_1} \cdots - 2\hat{e}_{i_{3-q}} )\,.
\end{equation}
\end{subequations}
It should be noticed that $v^*$ belongs to the lattice dual to that where $v$ lives.
An immediate consequence of the definition of Hodge duality here presented is that for any discrete form one has
\begin{equation}
\#\#\phi = \phi\,
\end{equation}
just as it happens in the  continuum framework. It should be underlined that there exist discretizations of the Hodge duality operation that do not "mimic" this property of the continuum.

We  also  introduce the coderivative of a discrete $q$-form, which we define as the $(q-1)$-form given by
\begin{equation}
(\delta \phi) = (-1)^q \,\#\, d \,\#\, \phi\,.
\end{equation}
A careful application of the above definition leads to the result
\begin{equation}
(\delta \phi)_{(q-1)\text{-cell}} = \sum_{\substack{q\text{-cell} /\\ (q-1)\text{-cell} \in \partial q\text{-cell}}} \phi_{q\text{-cell}}
\end{equation}
 where $\partial q$-cell is the boundary of the $q$-cell.
To obtain the components of the coderivative, we find it convenient to introduce the "overlined" partial derivative
\begin{equation}
\overline{\partial_a} f(v) \equiv f(v) - f(v-\mathbb{L}_a) = \partial_a f(v-\mathbb{L}_a).
\end{equation}
Then it can be shown that
\begin{equation}
(\delta \phi)_{i_1 \cdots i_{q-1}}(v) = - \partial_k\, \phi_{k\,i_1 \cdots i_{q-1}}(v-\mathbb{L}_k) = -\overline{\partial_k}\, \phi_{k\,i_1 \cdots i_{q-1}}(v)\, .
\end{equation}
Successive applications of the coderivative yield
\begin{equation}
\delta\delta \phi= 0\,.
\end{equation}
Next, we show how to discretize  integration by parts in a mimetic form. In the formula that follows, the summation is over all the  sites of the infinite cubic lattice. We have
\begin{equation*}
\sum_v \lambda_v \, \partial_a \mu_v = \sum_v \lambda_v (
\mu_{v+\mathbb{L}_a} - \mu_v) =  \sum_v (\lambda_{v-\mathbb{L}_a} -
\lambda_v) \mu_v = - \sum_v  (\lambda_v - \lambda_{v-\mathbb{L}_a})
\mu_v\,,
\end{equation*}
hence
\begin{equation}
\sum_v \lambda_v \, \partial_a \mu_v = - \sum_v  (\partial_a
\lambda_{v-\mathbb{L}_a}) \mu_v = - \sum_v  (\overline{\partial_a}
\lambda_v) \mu_v \,.
\end{equation}

It will also be useful the following definition. Given two $q$-forms in the lattice,  their inner product will be
\begin{equation}
\langle \phi, \psi \rangle = \langle \phi, \psi \rangle_{\mathbb{R}} + \langle \phi, \psi \rangle_{\mathbb{R}^d}
\end{equation}
where
\begin{align}
\langle \phi, \psi \rangle_{\mathbb{R}} &\equiv \sum_{\substack{q\text{-cell}>0 /\\ q\text{-cell}\in \mathbb{R}}}
\phi_{q\text{-cell}} \psi_{q\text{-cell}}= \langle \psi, \phi \rangle_{\mathbb{R}}
\\
\langle \phi, \psi \rangle_{\mathbb{R}^d} &\equiv \sum_{\substack{q\text{-cell}>0 /\\ q\text{-cell}\in \mathbb{R}^d}} \phi_{q\text{-cell}} \psi_{q\text{-cell}}= \langle \psi, \phi \rangle_{\mathbb{R}^d}.
\end{align}
Here, $q\text{-cell}>0$ means that the cells are taken  with just one of the two possible orientations. In components, the inner product is written as
\begin{equation}
\langle \phi, \phi' \rangle_{\mathbb{R}} =   \frac{1}{q!}\sum_{v \in \mathbb{R}} \phi_{i_1 \cdots i_q}(v)\, \phi'_{i_1 \cdots i_q}(v) \quad,\quad
\langle \phi, \phi' \rangle_{\mathbb{R}^d} =   \frac{1}{q!}\sum_{v \in \mathbb{R}^d} \phi_{i_1 \cdots i_q}(v)\, \phi'_{i_1 \cdots i_q}(v)
\end{equation}
and
\begin{equation}
\langle \phi, \phi' \rangle =   \frac{1}{q!}\sum_{v} \phi_{i_1 \cdots i_q}(v)\, \phi'_{i_1 \cdots i_q}(v) \,.
\end{equation}
The inner product satisfies the following properties
\begin{align}
\langle d\phi, \psi \rangle_{\mathbb{R}} &= \langle \phi, \delta\psi \rangle_{\mathbb{R}} \label{eq01-01}
\\
\langle \phi, \# \psi \rangle_{\mathbb{R}} &= \langle \# \phi, \psi \rangle_{\mathbb{R}^d}\label{eq01-02}\,.
\end{align}
that mimic well known properties of the continuum. From them it is immediate that
\begin{align}
\langle d\phi, \psi \rangle &= \langle \phi, \delta\psi \rangle
\\
\langle \phi, \# \psi \rangle &= \langle \# \phi, \psi \rangle \,.
\end{align}
To prove \eqref{eq01-01} let us take a  $q$-form $\varphi$ and a  $(q+1)$-form $\psi$. Then
\begin{align*}
\langle \phi, \delta\psi \rangle_{\mathbb{R}} &= \frac{1}{q!} \sum_{v \in \mathbb{R}} \phi_{i_1 \cdots i_q}(v) (\delta\psi)_{i_1 \cdots i_q}(v) = -\frac{1}{q!} \sum_{v \in \mathbb{R}} \phi_{i_1 \cdots i_q}(v) \overline{\partial_k} \psi_{k\, i_1 \cdots i_q}(v)
\\
&= \frac{1}{q!} \sum_{v \in \mathbb{R}} \partial_k \phi_{i_1 \cdots i_q}(v)  \psi_{k\, i_1 \cdots i_q}(v).
\end{align*}
But
\[
 \psi_{k\, i_1 \cdots i_q}(v) =  \frac{1}{(q+1)!}\delta_{j_1 \cdots j_{q+1}}^{k\, i_1 \cdots i_q}\psi_{j_1 \cdots j_{q+1}}(v) \quad\text{and}\quad (d\phi)_{j_1 \cdots j_{q+1}}(v) = \frac{1}{q!} \delta_{j_1 \cdots j_{q+1}}^{k\, i_1 \cdots i_q} \partial_k \phi_{i_1 \cdots i_q}(v)
\]
therefore
\begin{align*}
\langle \phi, \delta\psi \rangle_{\mathbb{R}} &= \frac{1}{q!} \frac{1}{(q+1)!}\sum_{v \in \mathbb{R}} \partial_k \phi_{i_1 \cdots i_q}(v) \delta_{j_1 \cdots j_{q+1}}^{k\, i_1 \cdots i_q}\psi_{j_1 \cdots j_{q+1}}(v)
\\
&= \frac{1}{(q+1)!}\sum_{v \in \mathbb{R}} (d\phi)_{j_1 \cdots j_{q+1}}(v) \psi_{j_1 \cdots j_{q+1}}(v) = \langle d\phi, \psi \rangle_{\mathbb{R}}\,.
\end{align*}
In turn, to prove \eqref{eq01-02} consider a $q$-form $\varphi$ and a $(3-q)$-form $\psi$. We have
\begin{align*}
\langle \# \phi, \psi \rangle_{\mathbb{R}^d} &= \frac{1}{(3-q)!}\sum_{v^* \in \mathbb{R}^d} (\# \phi)_{i_1 \cdots i_{3-q}}(v^*) \psi_{i_1 \cdots i_{3-q}}(v^*)
\\
&= \frac{1}{(3-q)!} \frac{1}{q!} \sum_{v \in \mathbb{R}} \varepsilon_{i_1 \cdots i_{3-q}\, j_1 \cdots j_q} \phi_{j_1 \cdots j_q}(v) \psi_{i_1 \cdots i_{3-q}}(v^*)
\end{align*}
with
\[
v^* = v + \frac{\mathbb{L}}{2} (\hat{e}_1 +\hat{e}_2 + \hat{e}_3 - 2\hat{e}_{i_1} \cdots - 2\hat{e}_{i_{3-q}}) = v - \frac{\mathbb{L}}{2} (\hat{e}_1 +\hat{e}_2 + \hat{e}_3 - 2\hat{e}_{j_1} \cdots - 2\hat{e}_{j_q})
\]
On the other hand
\[
\frac{1}{(3-q)!} \varepsilon_{i_1 \cdots i_{3-q}\, j_1 \cdots j_q} \psi_{i_1 \cdots i_{3-q}}(v^*) = \frac{1}{(3-q)!} \varepsilon_{j_1 \cdots j_q\, i_1 \cdots i_{3-q}} \psi_{i_1 \cdots i_{3-q}}(v^*) = (\# \psi)_{j_1 \cdots j_q}(v)
\]
hence we finally have
\[
\langle \# \phi, \psi \rangle_{\mathbb{R}^d} = \frac{1}{q!} \sum_{v \in \mathbb{R}} \phi_{j_1 \cdots j_q}(v) (\# \psi)_{j_1 \cdots j_q}(v) = \langle \phi, \# \psi \rangle_{\mathbb{R}}\,.
\]
To conclude this section we define the discrete versions of the most common differential operators of vectorial calculus: gradient, curl, divergence and Laplacian. Given a scalar field or 0-form $\lambda$, its  discrete gradient $\mathcal{G}\lambda$ is defined as
\begin{equation}
\mathcal{G}\lambda \equiv d \lambda \quad\text{hence}\quad (\mathcal{G}\lambda)_i(v) = \partial_i \lambda (v)\,.
\end{equation}
We define the discrete curl $\mathcal{R} A$ and divergence $\mathcal{D} A$ of a vector field or 1-form $A$ as follows.
\begin{align}
\mathcal{R} A &\equiv \# d A = \delta \# A, &&\text{hence ~~} (\mathcal{R} A)_i(v^*) = \frac{1}{2} \varepsilon_{i\,j\,k} (d A)_{j\,k}(v) = \varepsilon_{i\,j\,k} \partial_j A_k(v)
\\
\mathcal{D} A &\equiv -\delta A, &&\text{therefore ~~} (\mathcal{D} A)_v = \partial_i A_i(v-\mathbb{L}_i) =  \overline{\partial_i} A_i(v)
\end{align}
where $v^*=v + \mathbb{L}_{123}/2 - \mathbb{L}_i$. Since $d\,d=\delta\,\delta=0$ one has
\begin{equation}
\mathcal{R}\; \mathcal{G} = \mathcal{D}\; \mathcal{R} = 0 ,
\end{equation}
which mimics what occurs in the continuum framework.
We define the discrete Laplacian of a $q$-form $\phi$  as the  $q$-form given by
\begin{equation}
\nabla^2 \phi \equiv -(d \delta + \delta d) \phi ,
\end{equation}
and it can be shown that
 \begin{equation}
 (\nabla^2 \phi)_{i_1 \cdots i_q}(v) =  \overline{\partial_k}\partial_k \phi_{i_1 \cdots i_q}(v) = \partial_k\partial_k \phi_{i_1 \cdots i_q}(v - \mathbb{L}_k)\,.
\end{equation}

\section{ Abelian Chern-Simons Theory and the Gauss Linking Number}

Let us briefly review how the Gauss Linking Number can be obtained through the classical theory in the continuum \cite{Lorenzo-92,Lorenzo-02}. The action of the Abelian Chern-Simons theory, coupled to an external source, can be written down as

\begin{equation}
 S= \int d^{3}x (\frac{1}{2} \varepsilon^{ijk} A_{i}(x) \partial_{j} A_{k}(x)- \alpha A_{i}(x)J^{i}(x)) .
\end{equation}
It is metric independent; also, it is invariant under general coordinate transformations provided that the current $J^{i}(x)$ transforms as a contravariant vector density of weight $1$. The equation of motion is

\begin{equation}
  \varepsilon^{ijk}  \partial_{j} A_{k}(x) = \alpha J^{i}(x).
\end{equation}
For this equation to be consistent, the current must be conserved. This fact, as in Maxwell theory, is deeply related to gauge invariance. The equation of motion, being formally  Ampere's Law, is solved by the Biot-Savart law

\begin{equation}
 A_{i}(x) = \frac{\alpha}{4\pi} \int d^{3}x' \varepsilon^{ijk}  \frac{(x-x')^{j}}{|x-x'|^{3}} J^{k}(x') + \partial_{i}\Lambda\,\,  ,
\end{equation}
where $\Lambda$ is an arbitrary function that carries the gauge freedom mentioned above. If we take the current as the "path coordinate" of a closed path (i.e., a loop $C$)
\begin{equation}
 \chi^{i}(x)= \oint_{C} dx'^{i} \delta^{3}(x-x'),
 \end{equation}
substitute into the "Biot-Savart" expression and finally in the action, we obtain that the action on-shell yields

\begin{equation}\label{son}
 S_{on shell} = -\frac{\alpha^{2}}{8\pi} \oint_{C} dx^{i} \oint_{C} dx'^{j} \varepsilon^{ijk} \frac{(x-x')^{k}}{|x-x'|^{3}},
 \end{equation}
 which is $-\frac{\alpha^{2}}{2}$ times the Gauss Linking Number $L(C ,C')$, evaluated at $C' = C$ \cite{rolfsen}. Indeed, the above expression is ill defined, since the denominator vanishes whenever $x$ and $x'$ coincide. If we take two different (non-intersecting) curves instead of just one, this problem is cured. This is related to the issue of "framing" \cite{rolfsen}, which will be discussed latter. The Gauss Linking Number can also be written down in terms of line and surface integrals, which will be particularly useful for studying the lattice version of this link invariant. To this end, we find useful the following notation. Define Greek indexes as "composed indexes" $\mu = (i\,x)$, comprising a continuous and a discrete one. Also we introduce the following summation convention
\begin{equation}
H^\mu A_\mu \equiv \int d^3x\, B^{ix} A_{ix}\,.
\end{equation}
It is worth noticing that the the path coordinate (which may be defined for open paths as well) obeys
\begin{align}
\chi(\gamma_1\cdot\gamma_2) &=  \chi(\gamma_1) + \chi(\gamma_2),
\\
\chi(\bar{\gamma}) &=  -\chi(\gamma),
\end{align}
where $\gamma_1\cdot\gamma_2$ is the path product and  $\bar{\gamma}$ is the path opposite  to  $\gamma$.
Also, it verifies the differential constraint
\[
\partial_{ix} \chi^{ix}(\gamma) = \int_{\gamma} dy^i \partial_{ix}\delta(x-y)  = -\int_{\gamma} dy^i \partial_{iy}\delta(x-y) = \delta(x-x_{\bm{\cdot}\gamma}) - \delta(x-x_{\gamma\bm{\cdot}})\,.
\]
In particular if the path is a loop $C$ one has
\begin{equation}\label{eq02-01}
\partial_{ix} \chi^{ix}(C) = 0\,.
\end{equation}
Hence, there exists a "vector potential" $m_{ix}(C)$ of $\chi^{ix}(C)$ obeying
\begin{equation}
 \chi^{ix}(C) = \varepsilon^{i\,j\,k} \partial_{jx} m_{kx}(C)\,,
\end{equation}
which is defined up to the gradient of an arbitrary scalar field.
To calculate a convenient "vector potential" $m_{ix}(C)$, take  a surface $\Sigma_C$ whose boundary is the loop $C$. From Stokes theorem we have
 \begin{align*}
0 &= \oint_c d\vec{l}\cdot \vec{A} - \int_{\Sigma_C} d\vec{S} \cdot \vec{\nabla}\times\vec{A} = \oint_c dy^i A_{iy} - \int_{\Sigma_C} dS_k(y) \varepsilon^{k\,j\,i} \partial_{jy} A_{iy}
\\
&= \int d^3x A_{ix} \left[ \oint_c dy^i \delta(x-y) - \int_{\Sigma_C} dS_k(y) \varepsilon^{k\,j\,i} \partial_{jy} \delta(x-y) \right]
\\
&= \int d^3x A_{ix} \left[ \oint_c dy^i \delta(x-y) -  \varepsilon^{i\,j\,k} \partial_{jx} \int_{\Sigma_C} dS_k(y) \delta(x-y) \right] \; ,
\end{align*}
where $A_{ix}$ is any vector field. Hence we can write
\begin{equation}
\chi^{ix}(C) = \varepsilon^{i\,j\,k} \partial_{jx} M_{kx}(\Sigma_C) = \partial_{jx} \eta^{ij}_x(\Sigma_C) \, ,
\end{equation}
where the particular "vector potential" $M_{kx}(\Sigma_C)$ and its Hodge dual (or "co-potential") $\eta^{ij}_x(\Sigma_C)$ are given by
\begin{align}
M_{kx}(\Sigma_C) &= \int_{\Sigma_C} dS_k(y) \delta(x-y)
\\ \label{eq02-04}
\eta^{ij}_x(\Sigma_C) &=  \int_{\Sigma_C} \varepsilon^{i\,j\,k} dS_k(y) \delta(x-y)\,.
\end{align}
If $u$ and $v$ parameterize the surface $\Sigma_C$, and its  orientation is such that
\begin{equation}
d\vec{S} (y) = \frac{\partial \vec{y}}{\partial u} \times \frac{\partial \vec{y}}{\partial v} \,du \,dv  ,
\end{equation}
we shall have

\[
\varepsilon^{i\,j\,k} dS_k(y) = \varepsilon^{i\,j\,k} \varepsilon_{k\,n\,l}  \frac{\partial y^n}{\partial u} \frac{\partial y^l}{\partial v} \,du \,dv\, = \frac{\partial y^{[i}}{\partial u} \frac{y^{j]}}{\partial v} \,du \,dv .
\]
Hence, $M(\Sigma_C)$ y $\eta(\Sigma_C)$ may also be written as
\begin{align}
M_{kx}(\Sigma_C) &=  \frac{1}{2} \varepsilon_{i\,j\,k} \int_{\Sigma_C} d \Sigma_y^{i\,j}\,  \delta(x-y)
\\
\eta^{ij}_x(\Sigma_C) &=  \int_{\Sigma_C} d \Sigma_y^{i\,j}\, \delta(x-y)
\\
\end{align}
with
\begin{align}
d \Sigma_y^{i\,j} &\equiv \frac{\partial y^{[i}}{\partial u} \frac{y^{j]}}{\partial v} \,du \,dv\, \,.
\end{align}
In terms of the "vector potential" $M_{kx}(\Sigma_C)$ the action on-shell  can be cast in the form
\begin{equation}\label{sup}
 S_{on shell} = -\frac{\alpha^{2}}{2} \oint_{C} dx^{i} \int_{\Sigma_C} dS_i(y) \delta(x-y)  ,
 \end{equation}
which provides a simple geometric interpretation of the Gauss linking number: it counts how many times one of the curves cuts the surface bounded by the other one (in this case both curves are the same, but we could replace one of them by another one in the formula obtained above).

This discussion can be presented in the formalism of differential forms. Indeed, we shall translate these results to the lattice framework taking advantage of the mimetic differential-forms language introduced in the previous section. The Chern-Simons action coupled to a current, in the lattice, can be written  as
\begin{equation}\label{sdisc}
S_{CS} =  k_{1} < A, \# dA >\,+ k_{2} < A,j >\,= k_{1} \sum_{l>0} A_{l} \# dA_{l} \,+ k_{2} \sum_{l>0} A_{l}j_{l} \,\,  ,
\end{equation}
where $A$ is a discrete 1-form (the lattice Chern-Simons field) and  $j$ is the current. The discrete 3-space comprises both the direct and dual lattices, as discussed at the end of the first section.  Under lattice gauge transformations $A\rightarrow A + d\lambda$, where $\lambda$ is a 0-form,
the action becomes

\begin{align}
S' = &k_{1} (< A, \# dA >\,+ < d\lambda, \# dA >) + k_{2} (< A,j > + < d\lambda,j >)\\ \nonumber  =&S + k_{1} <\lambda , \delta\# dA >\,+ k_{2} < \lambda, \delta j >\,\, .
\end{align}
Since $\delta\# dA = \delta \delta \#A= 0$, we find that the action will be gauge invariant provided $\delta j=0$. This conservation law, in turn, implies that $j$ derives from a potential $m$ (or a co-potencial $\eta$, such that $\# m = \eta$ )
\begin{equation}
j = \delta \# m = \delta \eta .
\end{equation}
To study the dynamics of the system, we make  arbitrary and infinitesimal variations $\overline{\delta}A$ of the fields. Then, the action varies as

\begin{align}
\overline{\delta}S = k_{1} (< A, \# d\overline{\delta}A >\,+ < \overline{\delta}A, \# dA >\,) + k_{2} < \overline{\delta}A,j > .
\end{align}
The variational principle states that $\overline{\delta}S =0 $ for any $\overline{\delta}A$, hence we have
\begin{align}
k_{1} (\delta\# A + \# dA\,) + k_{2} j = 0 ,
\end{align}
and finally
\begin{align}\label{ecumov}
 \delta\# A = - \frac{k_{2}}{2 k_{1}} j .
\end{align}
To solve this equation, we recall that $j$ derives from a co-potencial $\eta$, which yields
\begin{align}\label{solecumov}
  A = - \frac{k_{2}}{2 k_{1}} \# \eta .
\end{align}
To proceed further, we must pick an appropriate choice for the external current $j$, i.e., one that mimics the loop coordinate $\chi^{ix}(C)$ of the continuum. Let us consider a path   $\gamma$ in the whole discrete space, that is, a succession of  links belonging to the union of the direct and dual lattices. Then we define the lattice path coordinate  $\chi(\gamma)$ as the 1-form given by

\begin{equation}
\chi_l(\gamma) = \sum_{l' \in \gamma} (\delta_{l,l'} - \delta_{\bar{l},l'})\,.
\end{equation}
$\chi_l(\gamma)$ counts how many times the path $\gamma$  passes along $l$ minus the number of times passing along $\bar{l} $. To explore further how far  $\chi_l(\gamma)$ mimics the former path coordinate let us define, as in the continuum, the contraction of two  1-forms $A$ and $H$ by
\begin{equation}
H_\mu A_\mu \equiv \sum_{l>0} H_l A_l= \sum_{v} H_i(v)\, A_i(v) = \langle H, A \rangle\, ,
\end{equation}
where the sums are over links and sites of $\mathbb{R}\cup\mathbb{R}^d$. The contraction of $\chi(\gamma)$ with the 1-form $A$ is
\[
\chi_\mu(\gamma) A_\mu = \sum_{l>0} \chi_l(\gamma) A_l = \sum_{l' \in \gamma} \; \sum_{l>0} A_l (\delta_{l,l'} - \delta_{\bar{l},l'}) =  \sum_{l' \in \gamma} \; \sum_{l>0} A_{l'} (\delta_{l,l'} + \delta_{l,\overline{l'}})
\]
hence we finally have
\begin{equation}
\chi_\mu(\gamma) A_\mu = \langle \chi(\gamma), A \rangle = \sum_{l' \in \gamma}  A_{l'}\,.
\end{equation}
As in the continuum, the path coordinate obeys
\begin{align}\label{eq02-05}
\chi(\gamma_1\cdot\gamma_2) &=  \chi(\gamma_1) + \chi(\gamma_2)
\\
\label{eq02-06}
\chi(\bar{\gamma}) &=  -\chi(\gamma_1)
\end{align}
Also, a careful calculation shows that the coderivative of $\chi(\gamma)$ is given by
\begin{equation}
(\delta\chi(\gamma))_v = -(\mathcal{D}\chi(\gamma))_v = \delta_{v,\gamma\bm{\cdot}}-\delta_{v,\bm{\cdot}\gamma} ,
\end{equation}
hence, for a loop $C$  one has
\begin{equation}
\delta\chi(C) = 0\,.
\end{equation}
 Therefore, $\chi(C)$ can be written as the coderivative of a 2-form $\eta(C)$ ("co-potential"), or the coderivative of the Hodge dual of a  1-form $m(C)$ ("potential") as follows
\begin{equation}
\chi(C) = \delta \eta(C) = \delta \# m(C)\,.
\end{equation}
It should be kept in mind that the space comprises both the direct and the dual lattices. Hence, when we talk about a loop we mean a "drawing" that may consist on several closed components, some of them in one of the lattices and some more in the other one.

The potential and copotential associated to $\chi(\gamma)$  can be explicitly constructed as follows. Let $\Sigma_C$ be a set of oriented plaquettes forming a surface bounded by the loop $C$. Define the 2-form $\eta(\Sigma_C)$ as
\begin{equation}
\eta(\Sigma_C)_p = \sum_{p'\in\Sigma_C} (\delta_{p',p} - \delta_{p',\bar{p}})\,.
\end{equation}
In words, $\eta(\Sigma_C)_p $ equals $1$ ($-1$) if $p\in\Sigma_C$  ($\bar{p}\in\Sigma_C$) , and vanishes in any other case. From $\eta(\Sigma_C)$ we construct the 1-form
\begin{equation}
M(\Sigma_C) \equiv \# \eta(\Sigma_C)\,.
\end{equation}
Notice that $M(\Sigma_C)$ equals $\pm 1$ when it is evaluated on the links   that "puncture"  the plaquettes composing $\Sigma_C$, depending on whether  these links are oriented as the punctured plaquette or in the opposite sense. In any other case $M(\Sigma_C)=0$.
It is not hard to see that
\begin{equation}
\chi(C) = \delta \eta(\Sigma_C)\, ,
\end{equation}
therefore, when one takes $\chi(C)$ as the external current of the lattice Chern-Simons theory, the solution \eqref{solecumov} of the field equation \eqref{ecumov} can be written down directly by replacing $\eta$ by $\eta(\Sigma_C)$. Then, substituting the solution \eqref{solecumov} into the action \eqref{sdisc} we can finally obtain the following expression for the action on shell

\begin{equation}
S_{on shell} = -\frac{k_{2}^{2}}{4k_{1}} \langle \#\eta(\Sigma_C),  \chi(C)\rangle = -\frac{k_{2}^{2}}{4k_{1}} \langle M(\Sigma_C),  \chi(C)\rangle\,\, .
\end{equation}
This expression defines the Gauss Linking number among two loops $C_{1}$, $C_{2}$
\begin{equation}\label{gn}
\varphi (C_{1},C_{2}) = \langle M(\Sigma_{C_{1}}),  \chi(C_{2})\rangle = \sum_{l>0} M_{l}(\Sigma_{C_{1}})  \chi_{l}(C_{2}).
\end{equation}
The right hand side of equation \eqref{gn} detects whether or not there are plaquettes belonging to $\Sigma_{C_{1}}$ whose dual link (or the opposite one) coincides with some link  of $C_{2}$. This amounts to say that $\varphi (C_{1},C_{2})$ detects if  $C_{2}$  punctures $\Sigma_{C_{1}}$, i.e., if $C_{2}$ links $C_{1}$. Notice that this linking is only possible among  closed components of $C_{1}$ and $C_{2}$ that belong to different lattices. If all the components of both loops belong to, say, the direct lattice, the Gauss Linking number will always vanish. However, there is a way to detect entanglement among loops in the same lattice, as we shall show  next. For every site in the direct (dual )lattice, there are eight sites of the dual (direct) one "surrounding" it. Therefore, given a drawing in one of the lattices, there are eight different copies of this drawing in the other lattice, which can obtained by elementary translations along the oblique eight directions pointing towards the surrounding sites. Then, given two loops we define their "smeared" Gauss Linking Number $\varphi' (C_{1},C_{2})$ as the average of the former linking numbers among one of the loops and each copy of the other loop which is obtained by translating it "slightly" in the eight possible forms  described above. In other words, we set
\begin{equation}
\varphi' (C_{1},C_{2}) = \frac{1}{8} \sum_{a=1}^{8} \varphi (T_{a} C_{1},C_{2}),
\end{equation}
where $T_{a} C$ represents the a'th translation of the loop $C$ (there are 8 such possible translations). It is not difficult to prove that
\begin{equation}
 \frac{1}{8} \sum_{a=1}^{8} \varphi (T_{a} C_{1},C_{2}) = \frac{1}{8} \sum_{a=1}^{8} \varphi ( C_{1},T_{a}^{-1} C_{2}) ,
\end{equation}
and that $\varphi' (C_{1},C_{2})$ is invariant under lattice translations and (discrete) rotations, as one should expect.

The smeared Gauss Linking Number $\varphi' (C_{1},C_{2})$ can be seen as a lattice implementation of the idea of "framing" in the continuum framework \cite{rolfsen}: to avoid ambiguities when studying the self-linking of a loop, one makes a slightly translated  copy of the loop that do not intersect with it, and calculates the Gauss Linking Number among the original and the "framed" loop. Of course, the result depends on the particular choice of framing. In our case, one could say that the smeared Gauss Linking Number $\varphi' (C_{1},C_{2})$  takes into account all the possibilities of framing in the lattice on the same footing. An interesting feature of this construction is that $\varphi' (C_{1},C_{2})$ yields well defined results even in the case of intersecting loops. Also, it is interesting to notice that in those cases, unlike the non-intersecting ones, the results are not integers. For instance, consider two  loops composed by just one closed piece lying in the direct lattice. If the loops intersect each other at one site forming a cross, the contribution  to $\varphi' (C_{1},C_{2})$ at that site will be $ \frac{1}{2}$. If the intersection is among a straight segment of one of the loops and a corner of the other one, the result will be $\frac{1}{4}$. This is a sensible manner to take into account those limiting cases of linking: one could say, for instance, that a crossing is "half" a linking.

\section{Metrics $\bm{g^{ix\,jy}}$ and $\bm{g_{ix\,jy}}$ in the continuum}

The expression for the Gauss Linking Number in the continuum \eqref{son} suggests the existence of an underlying structure that we shall explore in this section: the generalized metrics in the space of transverse functions. This study was initiated  in the continuum framework in reference \cite{dibartolo}. A vectorial density field  $T^{ix}$ is transverse if its divergence vanishes.  Define the transverse projector as
\begin{equation}
\delta^{ix}_{Tjy} \equiv \delta^{ix}_{jy}  + \partial_{jy}\phi^{ix}_y
\end{equation}
where $\phi ^{ix}_y$ is any function satisfying
\begin{equation}\label{eq02-08}
\partial_{ix}\phi^{ix}_y = \delta(x-y) \, ,
\end{equation}
and
\begin{equation}\label{eq02-08b}
\delta^{ix}_{jy} = \delta(x-y) \delta_{ij} \, .
\end{equation}
Since
\begin{equation}
\partial_{ix} \delta^{ix}_{Tjy} = 0\, ,
\end{equation}
$\delta^{ix}_{Tjy}$ projects a vector field into the space of transverse functions. Also, it acts as the identity operator in that space
\begin{equation}
\delta^{\mu}_{T\nu}\; T^\nu = T^\mu \quad\text{if }\; \partial_{ix} T^{ix} =0\,.
\end{equation}
There is a projector for each choice of $\phi$ (later, we shall refer to the relationship among different choices). Finally, we have
\[
\delta^{ix}_{T\mu} \; \delta^{\mu}_{Tjy} = \int d^3z \delta^{ix}_{Tkz} \; \delta^{kz}_{Tjy} =
\delta^{ix}_{Tjy} + \int d^3z \partial_{kz} \phi^{ix}_z \; \delta^{kz}_{Tjy}= \delta^{ix}_{Tjy},
\]
hence,  $\delta^{ix}_{Tjy}$ is indeed a projector.
We define the covariant $g_{\mu\,\nu}$ and contravariant $g^{\mu\,\nu}$ metrics in the transverse-forms space as

\begin{align}
g_{ix\,jy}  &=  \varepsilon_{ijk}\phi^{kx}_y \label{eq02-09} ,
\\[3mm]
g^{ix\,jy}  &\equiv -\varepsilon^{ijk} \partial_{kx} \delta(x-y) , \label{eq02-10}
\end{align}
where $\phi_T$ is asked to obey, besides \eqref{eq02-08}
\begin{align}
\phi^{iy}_x &= -\phi^{ix}_y \label{eq02-11}
\\
\phi^{ix}_{(y-z)} &= \phi^{i(x+z)}_y\,. \label{eq02-13}
\end{align}
These properties imply that
\begin{equation}
\partial_{kx}\phi^{ix}_y = -\partial_{ky}\phi^{ix}_y\, ,
\end{equation}
hence, the $g's$ are symmetric
\begin{align}
g_{ix\,jy}  &= g_{jy\,ix} \label{eq02-19} ,
\\[2mm]
g^{ix\,jy}  &= g^{jy\,ix} \label{eq02-18} ,
\end{align}
and their contraction yields

\begin{equation}
g^{ix\,\mu} \,g_{\mu\,jy} = \delta^{ix}_{jy}  + \partial_{jy}\phi ^{ix}_y \equiv \delta^{ix}_{Tjy}\,.
\end{equation}

The contravariant or inverse metric $g^{\mu\,\nu}$ metric obeys
\begin{equation}\label{eq02-20}
\partial_{ix} g^{ix\,jy} = \partial_{jy} g^{ix\,jy} = 0 ,
\end{equation}
and acting on a vector field  $A_u$ yields its curl
\begin{equation}\label{eq02-21}
g^{ix\,\mu}\, A_\mu = \varepsilon^{ijk}\partial_{jx} A_{kx} = (\mathcal{R} A)^ {ix}\,.
\end{equation}
If  $T^\mu$ is a transverse vectorial density, its potential vector  $A_\mu$ defined by
\begin{equation}
T^\mu = (\mathcal{R}A)^\mu
\quad\text{or}\quad
T = \delta \# A = \# dA \,  ,
\end{equation}
can be written down as
\begin{equation}
A_\mu = g_{\mu\,\nu}T^\nu .
\end{equation}
In fact
\begin{equation*}
(\mathcal{R}A)^\mu = g^{\mu\,\nu}A_\nu = g^{\mu\,\nu}g_{\nu\,\lambda}T^\lambda = \delta^{\mu}_{T\lambda} T^\lambda = T^\mu \,.
\end{equation*}
A choice of $\phi^{ix}_y$ determines a choice of metric and  a particular solution for the potential vector. Changing $\phi^{ix}_y$ amounts to making a gauge transformation $A' = A + d\lambda$.

Given two transverse vectorial densities  $T_1$ and $T_2$ we define their inner product as
\begin{equation}\label{eq02-14}
g(T_1,T_2) = T_1^{\mu} g_{\mu\,\nu} T_2^{\nu}\, ,
\end{equation}
which is independent of the choice of prescription for $\phi ^{ix}_y$, due to the transverse character of the $T's$.
The inner product \eqref{eq02-14} can also be written in terms of vector potentials. Given  $T_1^\mu = g^{\mu\,\alpha}A^1_{\alpha}$ and $T_2^\nu = g^{\nu\,\beta}A^2_{\beta}$ one has
\[
g(T_1,T_2) = T_1^{\mu} g_{\mu\,\nu} T_2^{\nu} = g^{\mu\,\alpha}A^1_{\alpha} \; g_{\mu\,\nu} \; g^{\nu\,\beta}A^2_{\beta} = A_{1\;\alpha} \; \delta^{\alpha}_{T\nu} g^{\nu\,\beta} \; A^2_{\beta} = A^1_{\alpha} \; g^{\alpha\,\beta} \;A^2_{\beta}\, .
\]
Taking $A \equiv A^1=A^2$ one obtains the Abelian Chern-Simons action.

Let us study some useful prescriptions. One of them, the "transverse prescription" is given by

\begin{equation}
\phi^{ix}_{0\,y} \equiv -\frac{1}{4\pi}\partial^{ix}\frac{1}{|x-y|} =
\frac{1}{4\pi}\frac{(x-y)^i}{|x-y|^3} .
\end{equation}
Since
\begin{equation}
\delta(x-y) = \nabla^2 \left(-\frac{1}{4\pi}\frac{1}{|x-y|}\right) ,
\end{equation}
it is immediate  to check \eqref{eq02-08}. Also, \eqref{eq02-11} and \eqref{eq02-13} are fulfilled. The covariant metric corresponding to  this prescription is given by
\begin{equation}
g_{0\,ix\,jy} = \frac{\varepsilon_{ijk}(x-y)^k}{4\pi|x-y|^3} =
\varepsilon_{ijk}\partial^{kx}\nabla^{-2}\delta(x-y) \, ,
\end{equation}
and it is transverse
\begin{equation}
\partial^{ix}g_{0\,ix\,jy} = \partial^{jy}g_{0\,ix\,jy} = 0 \,.
\end{equation}
Taking the loop coordinates $\chi^{\mu}$ as $T's$ and choosing the metric precisely in this prescription we see that the inner product $T_1^{\mu} g_{\mu\,\nu} T_2^{\nu}$ just corresponds to the Gauss Linking Number $\varphi (C_{1},C_{2})= \chi^{\mu}(C_{1}) g_{\mu \nu} \chi^{\nu}(C_{2})$. Since the result is independent of prescription, it could employed another ones to present different formulae for the Gauss invariant, which could be convenient in the lattice framework (the Green's function of the Laplacian that appears in the conventional formula in the continuum can not be put in a mimetic fashion).

The following prescription will be particularly useful in its lattice version. Let $h$ be a continuous path going from the origin to the spatial infinity.   $z(\tau)$  label the points of the path, $\tau$ being a parameter and  $z(0)=0$ the origin. We define the curves  $h_{y}$, obtained by translating  $h$ until its initial point coincides with the point $y$. Then, if  $z(\tau,h_{y})$ is the point of the curve  $h_{y}$ corresponding to the parameter value $\tau$, one has
\begin{equation}
z(\tau,h_{y}) = z(\tau) + y \,.
\end{equation}
It can be shown that the path coordinate $\chi^{kx}$ of the curves
$h_y$ satisfies
\begin{equation}\label{eq02-12}
\chi^{kx}(h_{(y-v)}) = \chi^{k(x+v)}(h_y)\,.
\end{equation}
From this construction, let us define the "bundle of paths" prescription as follows
\begin{equation}\label{eq02-29}
\phi^{ix}_{h~y} \equiv \frac{1}{2} \left[\int_{h_y} dz^i \delta(x-z) - \int_{h_x} dz^i \delta(y-z)\right] = \frac{1}{2} \left[\chi^{ix}(h_{y}) -  \chi^{iy}(h_{x}) \right] \,.
\end{equation}
It can be easily seen that $\phi^{ix}_{h~y}$ obeys  \eqref{eq02-11}. On the other hand, using \eqref{eq02-12} one has
\[
\phi^{ix}_{h~(y-\varpi)}  = \frac{1}{2} \left[\chi^{ix}(h_{y-\varpi}) -  \chi^{i(y-\varpi)}(h_{x}) \right]  = \frac{1}{2} \left[\chi^{i(x+\varpi)}(h_{y}) -  \chi^{iy}(h_{(x+\varpi)}) \right] = \phi^{i(x+\varpi)}_{h~y} ,
\]
hence \eqref{eq02-13} is also fulfilled. Regarding condition  \eqref{eq02-08}, one has
\begin{align*}
2\partial_{ix}\phi^{ix}_{h~y} &=
\int_{h_{y}} dz^i \partial_{ix}\delta(x-z) - \partial_{ix} \int_{h_x} dz^i \delta(y-z)
= -\int_{h_{y}} dz^i \partial_{iz}\delta(x-z) + \partial_{iy} \int_{h_x} dz^i \delta(y-z)
\\
&=-\int_{h_{y}} dz^i \partial_{iz}\delta(x-z) - \int_{h_x} dz^i \partial_{iz}\delta(y-z)
=-[-\delta(x-y)]-[-\delta(y-x)] = 2\delta(x-y)\,.
\end{align*}
The covariant metric associated to this prescription is then
\begin{align}
g_{h\,ix\,jy} &\equiv \varepsilon_{ijk}\phi^{kx}_{h~y}
= \frac{\varepsilon_{ijk}}{2}\left[\int_{h_y} dz^k \delta(x-z) - \int_{h_x} dz^k \delta(y-z)\right] \nonumber
\\
&=\frac{\varepsilon_{ijk}}{2}\left[\chi^{kx}(h_y) - \chi^{ky}(h_x)\right] \label{eq02-27}
\end{align}
Regardless which prescription for the metric we employ, it should be stressed that  the Gauss Linking number  can be written as
\begin{equation}\label{tal}
\varphi (C_{1},C_{2})= \chi^{\mu}(C_{1}) g_{\mu \nu} \chi^{\nu}(C_{2})\, .
\end{equation}
In the "bundle of paths" prescription, the interpretation of \eqref{tal} is as follows. The paths of the bundle that start in one of the loops  form a infinite tube that ends at the spatial infinity. Now, this tube plays the role of the surface ${\Sigma_{C}}$ of equation \eqref{sup}: the loops $C_{1}$ and $C_{2}$ are linked if one of them cuts the tube attached to the other one.

\section{Metrics $\bm{g^{\mu\,\nu}}$ y $\bm{g_{\mu\,\nu}}$ in the lattice}

Now we turn to discrete space. We shall say that a 1-form in the lattice is transverse if its coderivative (divergence) vanishes. A projector onto the vector space of 1-forms is given by
\begin{equation}
\delta^{iv}_{Tjw} \equiv \delta_{i,j} \delta(v-w)  + \partial_{jw}\phi^{iv}_{w}
\end{equation}
where $\phi^{iv}_w$ is a function that satisfies
\begin{equation}\label{eq02-07}
\overline{\partial}_{iv}\phi^{iv}_w = \delta(v-w) \,.
\end{equation}

The following properties ensure that $\delta^{iv}_{Tjw}$ is a projector as claimed
\begin{align}
\overline{\partial}_{iv}\; \delta^{iv}_{Tjw} &= 0 \label{eq02-15}
\\
\delta^{\alpha}_{T\mu} \; \delta^{\mu}_{T\beta} &=  \delta^{\alpha}_{T\beta} \label{eq02-16}
\\
\delta^{\alpha}_{T\mu} \; T^\mu &= T^\alpha ~~ \forall\; T /\;  \delta T = 0\,. \label{eq02-17}
\end{align}
Let us prove \eqref{eq02-15}:
\[
\overline{\partial}_{iv} \delta^{iv}_{Tjw} = \overline{\partial}_{jv}\delta(v-w) + \partial_{jw}\overline{\partial}_{iv}\phi^{iv}_{w} = - \partial_{jw}\delta(v-w) + \partial_{jw}\delta(v-w) = 0\,.
\]
Regarding \eqref{eq02-16}, one has

\begin{align*}
\delta^{iv}_{T\mu}  \delta^{\mu}_{Tjw} &= \sum_{v"} \delta^{iv}_{Tcv"}  \; \delta^{cv"}_{Tjw} = \sum_{v"} \left [ \delta_{i,c} \delta(v-v")  + \partial_{cv"}\phi^{iv}_{v"}\right] \delta^{cv"}_{Tjw}
\\
&= \delta^{iv}_{Tjw} - \sum_{v"} \phi^{iv}_{v"}\; \overline{\partial}_{cv"} \delta^{cv"}_{Tjw} = \delta^{iv}_{Tjw} \,.
\end{align*}
Finally we prove \eqref{eq02-17}:
\[
\delta^{iv}_{T\mu} T^\mu = \sum_{w} \left [ \delta_{i,c} \delta(v-w)  + \partial_{cw}\phi^{iv}_{w}\right]  T^{cw} = T^{iv} - \sum_{w} \phi^{iv}_{w}\; \overline{\partial}_{cw}  T^{cw} = T^{iv} \,.
\]
The covariant metric in the lattice is a discretization of \eqref{eq02-10} defined as
\begin{equation}
g^{iv\,jw}  \equiv -\varepsilon_{ijk} \partial_{kv} \delta(v-w+\mathbb{L}_i - \frac{1}{2} \mathbb{L}_{123})\,.
\end{equation}
The following properties of $g^{iv\,jw}$ are "mimetizations" of  \eqref{eq02-21},
\eqref{eq02-20}, and \eqref{eq02-18} respectively
\begin{align}
g^{iv\,\mu} A_\mu &= (\mathcal{R} A)^{iv} \label{eq02-22}
\\
\overline{\partial}_{iv} \; g^{iv\,jw} &=0 \label{eq02-23}
\\
g^{\mu\,\nu} &= g^{\nu \, \mu}\,. \label{eq02-24}
\end{align}
Let us prove \eqref{eq02-22}, which states that the contraction of the metric with a vector  yields its curl
\[
g^{iv\,\mu} A_\mu = \sum_w g^{iv\,jw} A_{jw} = -\varepsilon_{ijk} \partial_{kv} A_{j(v+\mathbb{L}_i - \frac{1}{2} \mathbb{L}_{123})} = \varepsilon_{iab} \partial_{av} A_{b(v+\mathbb{L}_i - \frac{1}{2} \mathbb{L}_{123})} = (\mathcal{R} A)^{iv} \,.
\]
Next we prove that the metric is transverse  (property \eqref{eq02-23})
\[
 \overline{\partial}_{iv}\; g^{iv\,jw} = -\varepsilon_{ijk} \partial_{kv} \overline{\partial}_{iv} \delta(v-w+\mathbb{L}_i - \frac{1}{2} \mathbb{L}_{123}) = -\varepsilon_{ijk} \partial_{kv} \partial_{iv} \delta(v-w - \frac{1}{2} \mathbb{L}_{123}) = 0\,.
\]
Also, the metric is symmetric (property \eqref{eq02-24})
\begin{align*}
g^{jw\,iv} &= -\varepsilon_{jik}\; \partial_{kw} \delta(w-v+_j - \frac{1}{2} \mathbb{L}_{123})
= -\varepsilon_{ijk}\; \overline{\partial}_{kv}\; \delta(w-v+\mathbb{L}_j - \frac{1}{2} \mathbb{L}_{123})
\\
&= -\varepsilon_{ijk}\; \partial_{kv}\; \delta(w-v+\mathbb{L}_k+\mathbb{L}_j - \frac{1}{2} \mathbb{L}_{123}) = -\varepsilon_{ijk}\; \partial_{kv}\; \delta(w-v-\mathbb{L}_i + \frac{1}{2} \mathbb{L}_{123}) = g^{iv\,jw}\, .
\end{align*}

We take as the discretization of the covariant metric \eqref{eq02-09} the expression
\begin{equation}
g_{iv\,jw}  \equiv \varepsilon_{ijk}\;  \phi^{k(v+\frac{1}{2}\mathbb{L}_{123}-\mathbb{L}_j-\mathbb{L}_k)}_{w} = \varepsilon_{ijk}\;  \phi^{k(v+\mathbb{L}_i-\frac{1}{2}\mathbb{L}_{123})}_{w}.
\end{equation}
In order to be a satisfactory covariant metric in the transverse space we ask $\phi^{iv}_{w}$ to obey the discrete versions of equations
 \eqref{eq02-11} y \eqref{eq02-13}, that is
\begin{align}
\phi^{iw}_v &= -\phi^{i(v-\mathbb{L}_i)}_w  \quad\text{(without summation in $i$)}
\\
\phi^{iv}_{(w-\varpi)} &= \phi^{i(v+\varpi)}_w\,.
\end{align}
The last equations implies
\begin{align}
\overline{\partial}_{jv}\; \phi^{iv}_w &= - \partial_{jw} \phi^{iv}_w
\\
\partial_{jv} \phi^{iv}_w &= - \overline{\partial}_{jw}\; \phi^{iv}_w\, ,
\end{align}
as can be easily demonstrated. As in the continuum, the covariant metric satisfies the properties
\begin{align}
g^{\alpha\,\mu} \,g_{\mu\,\beta} &=  \delta^{\alpha}_{T\beta}\label{eq02-25}
\\
g_{\alpha\,\beta}  &= g_{\beta\,\alpha}\, , \label{eq02-26}
\end{align}
whose demonstration follows. Regarding  \eqref{eq02-25} one has
\begin{align*}
g^{iv\,\mu} \,g_{\mu\,jw} &= \sum_{v"} g^{iv\,a v"} \,g_{a v"\,jw} = -\varepsilon_{iak} \partial_{kv} \,g_{a (v+\mathbb{L}_i - \frac{1}{2} \mathbb{L}_{123})\,jw} = \varepsilon_{aik}\varepsilon_{ajb}\; \partial_{kv} \phi^{b(v+\mathbb{L}_i -\mathbb{L}_j-\mathbb{L}_b)}_{w}
\\
&= \delta_{i\,j} \partial_{kv} \phi^{k(v-\mathbb{L}_k)}_{w} - \partial_{jv} \phi^{i(v -\mathbb{L}_j)}_{w} = \delta_{i\,j} \overline{\partial}_{kv} \phi^{kv}_{w} - \overline{\partial}_{jv} \phi^{iv}_{w} = \delta_{i\,j} \delta(v-w) + \partial_{jw} \phi^{iv}_{w} = \delta^{iv}j{w}\,.
\end{align*}
In turn, \eqref{eq02-26} can be established as follows
\[
g_{jw\,iv} = \varepsilon_{jik}\;  \phi^{k(w+\mathbb{L}_j-\frac{1}{2}\mathbb{L}_{123})}_{v} = - \varepsilon_{jik}\;  \phi^{k(v-\mathbb{L}_k)}_{(w+\mathbb{L}_j-\frac{1}{2}\mathbb{L}_{123})} = \varepsilon_{ijk}\;  \phi^{k(v-\mathbb{L}_k-\mathbb{L}_j+\frac{1}{2}\mathbb{L}_{123})}_{w} = g_{iv\,jw} .
\]

As it happens in the continuum setting, the discrete metrics allow to find the "potential vector" $A_\mu$ whose curl is a given transverse vector field $T^\mu$
\begin{equation}
T^\mu = (\delta \# A)^\mu = (\# dA)^\mu = (\mathcal{R}A)^\mu\,.
\end{equation}
A solution to this equation is given by
\begin{equation}
A_\mu = g_{\mu\,\nu}T^\nu ,
\end{equation}
since
\begin{equation*}
(\mathcal{R}A)^\mu = g^{\mu\,\nu}A_\nu = g^{\mu\,\nu}g_{\nu\,\lambda}T^\lambda = \delta^{\mu}_{T\lambda} T^\lambda = T^\mu \,.
\end{equation*}

Next we shall study the discretization of the metric $g_{h\,\mu\,\nu}$ associated to the "bundle of paths" prescription defined by equation \eqref{eq02-27}. Let $h$ be a continuous path in one of the lattices (say the direct one) starting at the origin and ending at the spatial infinity.  For each site $v$  of both lattices we define, as before, a path $h_v$  obtained through rigid translations of $h$ in such a way that the new starting point be $v$. It can be seen that the lattice loop coordinate of this "bundle of paths" obeys
\begin{equation}\label{eq02-30}
\chi^{i v}(h_{w-\varpi}) = \chi^{i (v+\varpi)}(h_{w}) ,
\end{equation}
which is the discrete version of \eqref{eq02-12}.

The differential constraint for $\chi^{i v}(h_{w}) $ takes the form
\begin{equation}\label{eq02-32}
\overline{\partial}_{iv} \chi^{i v}(h_{w}) = \delta(v - w)
\end{equation}
which can be cast as
\[
\delta(v - w) = \overline{\partial}_{iv} \chi^{i v}(h_{w}) = \sum_i \left[\chi^{i v}(h_{w}) -  \chi^{i (v-\mathbb{L}_i)}(h_{w}) \right] = \sum_i \left[\chi^{i v}(h_{w}) -  \chi^{i v}(h_{(w+\mathbb{L}_i)}) \right]
\]
in virtue of \eqref{eq02-30}. From this, we have
\begin{equation}\label{eq02-33}
\partial_{iw} \, \chi^{i v}(h_{w}) = -\delta(v - w)\,.
\end{equation}

Now we discretize $\phi^{ax}_{h~y}$, given by equation \eqref{eq02-29} in the form
\begin{equation}\label{fid}
\phi^{iv}_{h~w} \equiv  \frac{1}{2} \chi^{iv}(h_w) -  \frac{1}{2}\chi^{iw}(h_{v+\mathbb{L}_i}) ,
\end{equation}
and define de covariant metric as
\begin{equation}\label{gcov}
g_{h~iv\,jw}  \equiv  \varepsilon_{ijk}\;  \phi^{k(v+\mathbb{L}_i-\frac{1}{2}\mathbb{L}_{123})}_{h~w}\,.
\end{equation}

$\phi^{iv}_{h~w}$ satisfies certain properties that make $g_{h~\mu\,\mu}$ to be a covariant metric metric. They are
\begin{align}
\overline{\partial}_{iv}\phi^{iv}_{h~w} &= \delta(v-w) \label{eq02-31}
\\
\phi^{iw}_{h~v} &= -\phi^{i(v-\mathbb{L}_i)}_{h~w} \label{eq02-34}
\\
\phi^{iv}_{h~(w-\varpi)} &= \phi^{i(v+\varpi)}_{h~w}\,.\label{eq02-35}
\end{align}

To establish \eqref{eq02-31} we use the differential constraint as given by  \eqref{eq02-32} and \eqref{eq02-33}
\[
\overline{\partial}_{iv}\phi^{iv}_{h~w} = \frac{1}{2} \overline{\partial}_{iv}\chi^{iv}(h_w) -  \frac{1}{2} \overline{\partial}_{iv} \chi^{iw}(h_{v+\mathbb{L}_i}) = \frac{1}{2} \delta(v-w) - \partial_{iv} \chi^{iw}(h_{v}) =  \delta(v-w)\,.
\]

The proofs  of \eqref{eq02-34} and \eqref{eq02-35} use  identity \eqref{eq02-30}
\[
\phi^{iw}_{h~v} = \frac{1}{2} \chi^{iw}(h_v) -  \frac{1}{2}\chi^{iv}(h_{w+\mathbb{L}_i}) =  \frac{1}{2} \chi^{iw}(h_v) -  \frac{1}{2}\chi^{i(v-\mathbb{L}_i)}(h_{w}) = -\phi^{i(v-\mathbb{L}_i)}_{h~w}\, ,
\]
as well as
\[
\phi^{iv}_{h~(w-\varpi)} = \frac{1}{2} \chi^{iv}(h_{w-\varpi}) -  \frac{1}{2}\chi^{i(w-\varpi)}(h_{v+\mathbb{L}_i}) = \frac{1}{2} \chi^{i(v+\varpi)}(h_{w}) -  \frac{1}{2}\chi^{iw}(h_{v+\varpi+\mathbb{L}_i}) = \phi^{i(v+\varpi)}_{h~w}\,.
\]
Hence, we have constructed a covariant  metric that mimics, in discrete space, the "bundle of paths" metric of the continuum. We write down a final expression for it, that should be compared with \eqref{eq02-27}, and that is obtained by substituting \eqref{fid} into \eqref{gcov}

\begin{equation}\label{gfinal}
g_{h~iv\,jw}= \frac{1}{2}  \varepsilon_{ijk} (\chi^{k(v+\mathbb{L}_i - \frac{1}{2} \mathbb{L}_{123})}(h_{w}) -  \chi^{k(w-\mathbb{L}_k)}(h_{v+\mathbb{L}_i - \frac{1}{2} \mathbb{L}_{123}}) )\, .
\end{equation}
 
\section{Summary and outlook}
As we saw, the Gauss Invariant $\varphi(C_1, C_2)$ can be written as 
\begin{equation}\label{NucleoGauss}
\varphi(C_1, C_2) = \chi^\mu(C_1) g_{\mu\nu} \chi^\nu(C_2)\,  ,
\end{equation}
where $g_{\mu\nu}$ is the covariant metric in any prescription. In turn, the Chern-Simons term can be expressed as 
\begin{equation}\label{NucleoCS}
S_{CS} = A_\mu g^{\mu\nu} A_\nu\,.
\end{equation}
Hence, the covariant and contravariant metrics are the kernels of the Gauss Invariant and of the Chern-Simons term, respectively, both in the continuum and in the lattice. This is not the only reason why these metrics are interesting. Indeed, they can be seen as building blocks for other knot invariants that are interesting in quantum gravity, as we shall briefly discuss. The key object to take into account is the Wilson operator $W_A(\gamma)$, which is the trace of the path ordered exponential of the non Abelian Chern-Simons potential
\begin{equation}
W_A(\gamma) = \text{Tr} \left[ P \; \text{exp} \oint_\gamma A_a dy^a\right]\,.
\end{equation}
Witten showed that there is a correspondence between the vacuum expectation value $<W(\gamma)>$ in the non Abelian Chern-Simons theory and the Jones and Homfly polynomials \cite{Witten}. Guadagnini, Martellini and Mintchev \cite{Guadagnini} found that the perturbative expansion of $<W(\gamma)>$ provides a systematic method to obtain analytical expressions of knot and link invariants. This perturbative expansion has the form
\begin{equation}
<W(\gamma)> = \int dA \; e^{iS_{CS}} W_A(\gamma) = N + N\sum_{n=1}^\infty \left(\frac{2\pi i}{\kappa}\right)^n <W(\gamma)>^{(n)}\, ,
\end{equation}
where the gauge potential $A$ belongs to the SU(N) algebra and $\kappa$ is a coupling constant that serves as perturbative parameter.
Analytical expressions for the first terms in the expansion are \cite{Guadagnini,Gae-93} 
\begin{align}
<W(\gamma)>^{(1)} &= -\frac{N^2-1}{2N} \varphi(\gamma)
\\
<W(\gamma)>^{(2)} &= -\frac{N^2-1}{2} \rho(\gamma) + \frac{1}{2}\left( <W(\gamma)>^{(1)} \right)^2
\\
<W(\gamma)>^{(3)} &= -\frac{N(N^2-1)}{2} \tau(\gamma) +<W(\gamma)>^{(1)}<W(\gamma)>^{(2)}   - \frac{1}{3}\left( <W(\gamma)>^{(1)} \right)^3\, ,
\end{align}
 where $\varphi(\gamma)$, $\rho(\gamma)$ and $\tau(\gamma)$ are knot invariants. The firs one is the self linking Gauss number, while the other are related with the second coefficient of the Alexander-Conway polynomial and the third coefficient of the Jones polynomial respectively. In terms of loop coordinates of higher rank \cite{Cayetano}, these objects are written as \cite{Gae-93}
\begin{align}
\varphi(\gamma) &= \chi^\mu g_{\mu\nu} \chi^\nu = 2 g_{\mu_1\mu_2} \chi^{\mu_1\mu_2}\,,
\\
\rho(\gamma) &= 2(-h_{\mu_1\mu_2\mu_3}\chi^{\mu_1\mu_2\mu_3}+ g_{\mu_1\mu_3}  g_{\mu_2\mu_4} \chi^{\mu_1\mu_2\mu_3\mu_4})\,,
\\
\tau(\gamma) &= 2\left( h_{\mu_1\mu_4\alpha}g^{\alpha\beta}h_{\mu_2\mu_3\beta} - h_{\mu_1\mu_2\alpha}g^{\alpha\beta}h_{\mu_3\mu_4\beta} \right)\chi^{\mu_1\mu_2\mu_3\mu_4}
-2g_{(\mu_1\mu_3}h_{\mu_2\mu_4\mu_5)_c}\chi^{\mu_1\mu_2\mu_3\mu_4\mu_5}
\nonumber
\\
&+\left(4g_{\mu_1\mu_4}g_{\mu_2\mu_5}g_{\mu_3\mu_6}  + g_{(\mu_1\mu_3}g_{\mu_2\mu_5}g_{\mu_4\mu_6)_c} \right)\chi^{\mu_1\mu_2\mu_3\mu_4\mu_5\mu_6}\,,
\end{align}
where the metrics are taken in the transverse prescription and we have set
\begin{equation}
h_{\mu_1\mu_2\mu_3} \equiv \varepsilon^{\alpha\beta\gamma}g_{\mu_1\alpha}g_{\mu_2\beta}g_{\mu_3\gamma} \quad\text{and}\quad \varepsilon^{ax\,by\,cz} \equiv \int d^3t \varepsilon^{abc} \delta(x-t)\delta(y-t)\delta(z-t)\,.
\end{equation}
These expressions show the important role that the metric plays in the construction of analytical formulae for the invariants, as claimed. These invariants are candidates to be quantum states of Loop Quantum Gravity, and it could be useful to have mimetic discrete versions of them in order to search for lattice solutions of quantum gravity. A first step in this program is the study of the metrics in the space of transverse vector densities that we have presented in this article.

\end{document}